\documentclass[aps,prb,twocolumn,letterpaper,superscriptaddress,showpacs]{revtex4-2}
\usepackage{graphicx}
\usepackage{CJK}
\usepackage{verbatim}
\usepackage{physics}
\usepackage[colorlinks,linkcolor=blue,anchorcolor=blue, citecolor=blue,urlcolor=blue,]{hyperref}
\usepackage{lineno}
\usepackage{bm}
\usepackage{mathptmx}
\makeatletter
\newcommand{\parallelsum}{\mathbin{\!/\mkern-5mu/\!}}
\begin{document}
%\linenumbers
\begin{CJK*}{UTF8}{bsmi}
%\title{Possible tunable superconductivity in Ni-based superconductor}
\title{Proposal to improve Ni-based superconductors via enhanced charge transfer}
\author{Zi-Jian Lang (\CJKfamily{gbsn}郎子健)}
\affiliation{Tsung-Dao Lee Institute \& School of Physics and Astronomy, Shanghai Jiao Tong University, Shanghai 200240, China}
\author{Ruoshi Jiang (\CJKfamily{gbsn}姜若诗)}
\affiliation{Tsung-Dao Lee Institute \& School of Physics and Astronomy, Shanghai Jiao Tong University, Shanghai 200240, China}
\author{Wei Ku (\CJKfamily{bsmi}顧威)}
\altaffiliation{corresponding email: weiku@sjtu.edu.cn}
%\email{weiku@sjtu.edu.cn}
\affiliation{Tsung-Dao Lee Institute \& School of Physics and Astronomy, Shanghai Jiao Tong University, Shanghai 200240, China}
\affiliation{Key Laboratory of Artificial Structures and Quantum Control (Ministry of Education), Shanghai 200240, China} 
\date{\today}
\begin{abstract}
Recently discovered superconductivity in hole-doped nickelate Nd$_{0.8}$Sr$_{0.2}$NiO$_2$ has attracted intensive attention in the field.
An immediate question is how to improve its superconducting properties.
Guided by the key characteristics of electronic structures of the cuprates and the nickelates, we propose that nickel chalcogenides with a similar lattice structure should be a promising family of materials.
Using NdNiS$_2$ as an example, through first-principle structural optimization and phonon calculation, we find this particular crystal structure a stable one.
We justify our proposal by comparing with CaCuO$_2$ and NdNiO$_2$ with regard to strength of the charge-transfer characteristics and the trend in their low-energy many-body effective Hamiltonians of doped hole carriers.
This analysis indicates that nickel chalcogenides host low-energy physics closer to that of the cuprates, with stronger magnetic interaction than the nickelates, and thus they deserve further experimental exploration.
Our proposal also opens up the possibility of a wide range of parameter tuning through ligand substitution among chalcogenides, to further improve superconducting properties.
\end{abstract}
\maketitle
\end{CJK*}

Hole doped nickelates, Re$_{1-x}$Sr$_x$NiO$_2$~\cite{Li,Li2020,Osada2021}, as Ni-based high-temperature superconductors, have attracted great attention in condensed matter physics recently.
They display type-\uppercase\expandafter{\romannumeral2} superconductivity, a dome-shaped superconducting phase~\cite{Li2020} and strange metal (linear resistivity) behavior in its normal state~\cite{Li}.
All of these characteristics suggest that these materials represent yet another family of unconventional superconductors.
Meanwhile, their strong temperature and doping dependent hall coefficient~\cite{Li2020}, negative magneto-resistance~\cite{Wen2020}, absence of long-range magnetic order~\cite{Hayward2003,Sawatzky2019} in the parent compound, and increasing normal-state resistivity in the overdoped regime~\cite{Li2020,Osada2020} also indicate rich underlying physics in this family of superconductors that might be absent in the cuprates~\cite{Ando2004,Ando2004res,Bozovic,Dagotto}.
Obviously, these nickelate superconductors belong to a promising family that could help unravel the longstanding puzzles of high-temperature superconductivity and even to find higher transition temperature $T_c$ beyond the cuprates.

So far within limited attempts, the highest $T_c$ of nickelates is only about 12K~\cite{Li2020,Osada2020}, one order of magnitude lower than that of the best cuprates~\cite{Schilling1993,Dagotto}.
Furthermore, at present, superconductivity is only found in thin films but not in bulk samples~\cite{Wen2020}, for reasons yet to be understood.
Significant experimental progress is thus expected upon improvement of sample quality.
On the other hand, it is of equal importance to seek other approaches to improve the superconducting properties besides the sample quality. 

Here we address this timely issue by first comparing the high-energy electronic structure of the cuprates and nickelates to identify their key characteristics to be the strength of charge transfer.
Based on this, we propose an alternative family of material, nickel chalcogenides, as a promising candidate to improve the superconducting properties.
Taking NdNiS$_2$ as an example, through density functional structure optimization and phonon calculation, we first demonstrate that this compound is stable under the same crystal structure as NdNiO$_2$.
The corresponding high-energy electronic structure confirms our expectation of an enhanced charge-transfer characteristic.
Furthermore, our local many-body diagonalization gives a ground state similar to those of the cuprates and nickelates, namely, a spin-singlet state with doped holes mostly residing in ligand $p$ orbitals~\cite{Lang2020}.
As anticipated, the corresponding effective electron-volt-scale Hamiltonian of hole carriers contains stronger spin interactions than NdNiO$_2$, suggesting a higher temperature scale in the phase diagrams.
Our study indicates that nickel chalcogenides are promising candidates for improved superconducting properties, and ligand substitution, e.g., NdNiS$_{2-x}$O$_x$ and NdNiS$_{2-x}$Se$_x$, would introduce a great deal of tunability for future experimental exploration.
%\vspace{-0.5cm}
\begin{figure}[b]
	\begin{center}
    \includegraphics[width=\columnwidth]{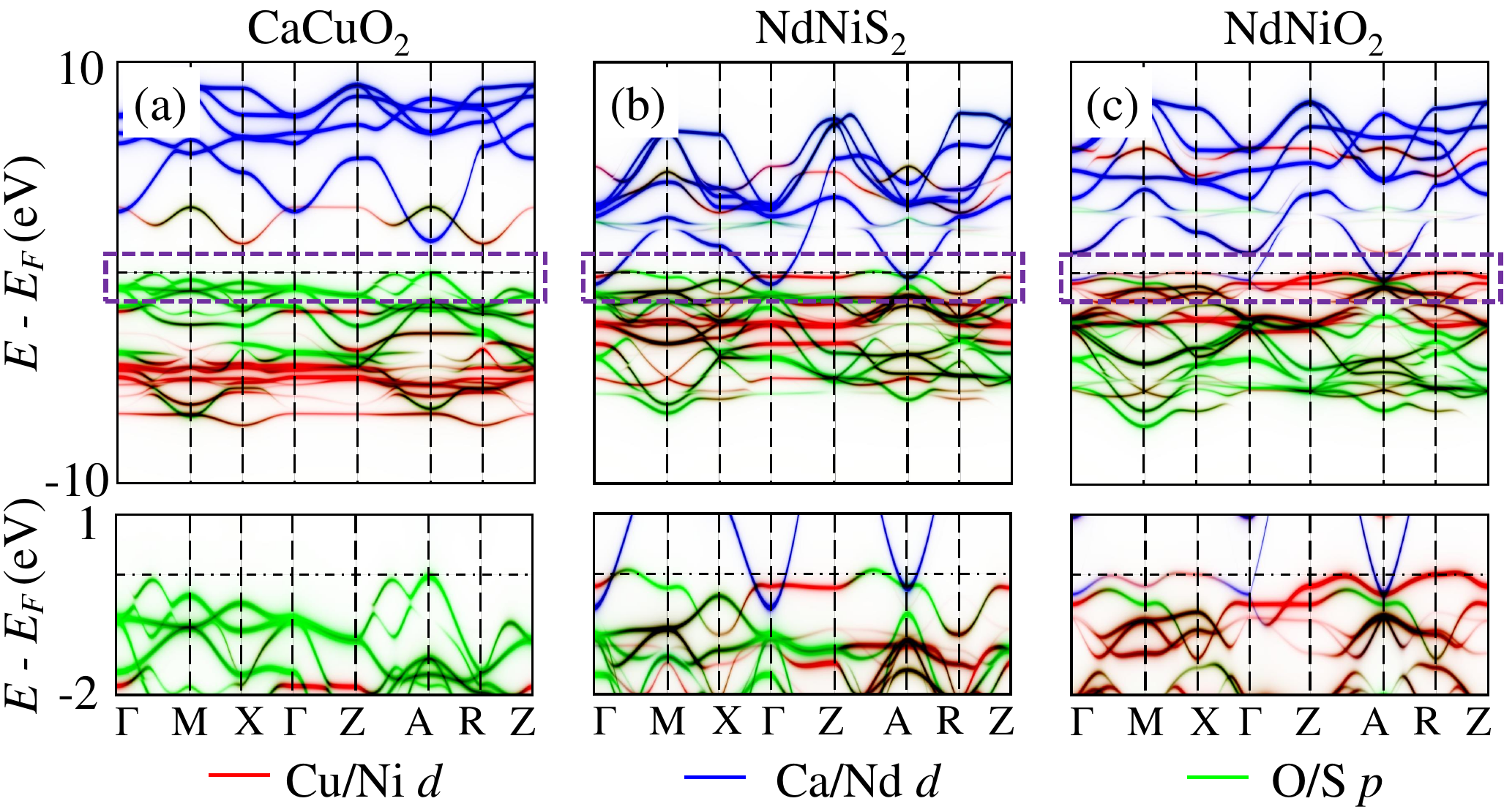}
	\end{center}
	\vspace{-0.6cm}
	\caption{Comparison of LDA+$U$ band structures of CaCuO$_2$, NdNiS$_2$ and NdNiO$_2$ under AFM order, unfolded in the one-Cu/Ni Brillouin zone. The red, blue, and green colors represent the weights of Cu/Ni, Ca/Nd, and O/S orbitals, and Nd $f$-orbitals are set by completely transparent. The lower panels show the magnified band structure of the purple dashed boxes in the upper panels. Notice the trend in the relative energies of O/S and Cu/Ni orbitals.}
	\label{bs}
	%\vspace{-0.6cm}
\end{figure}

To identify the key difference between the cuprates and the nickelates, we compare their high-energy electronic structure using density functional theory (DFT).
Since both the cuprates and the nickelates host strong antiferromagnetic (AFM) correlation~\cite{Lee2006,Cui2021,Lin2021} inherited from the unfrustrated square lattice of the spin-1/2 local moment~\cite{Anderson1950}, we calculate the band structures under AFM order within the LDA+$U$ approximation~\cite{Anisimov,Liechtenstein,Blaha2020,Kopnin1997,supp1} and unfold them to the one-Ni unit cell for a simpler visualization~\cite{Wei2010}.
Figures~\ref{bs}(a) and (c) show that compared with NdNiO$_2$, the main difference of CaCuO$_2$ at the large energy scale is the much lower energy of its $d$ orbitals (in red) relative to the O $p$ orbitals (in green), reflecting a much stronger charge-transfer nature well known to the community~\cite{Zannen,Zhang2020,Lechermann2020, Botana2020,Karp2020,Goodge2021,Jiang2020}.
Given that both families are doped spin-1/2 systems, it is reasonable to expect that promoting such a charge transfer characteristic should improve significantly the superconducting properties, due to various considerations of low-energy physics such as enhanced super-exchange interaction~\cite{Anderson1950,Weber2012,Lin2021} and renormalized kinetic energy.
Since there is no chemical way to further lower the orbital energy of Ni (other than replacing it by Cu), we are left with no choice but to raise the energy of the ligand $p$ orbitals, for example, by substituting O with S or Se.

\begin{figure}
	\begin{center}
		\includegraphics[width=\columnwidth]{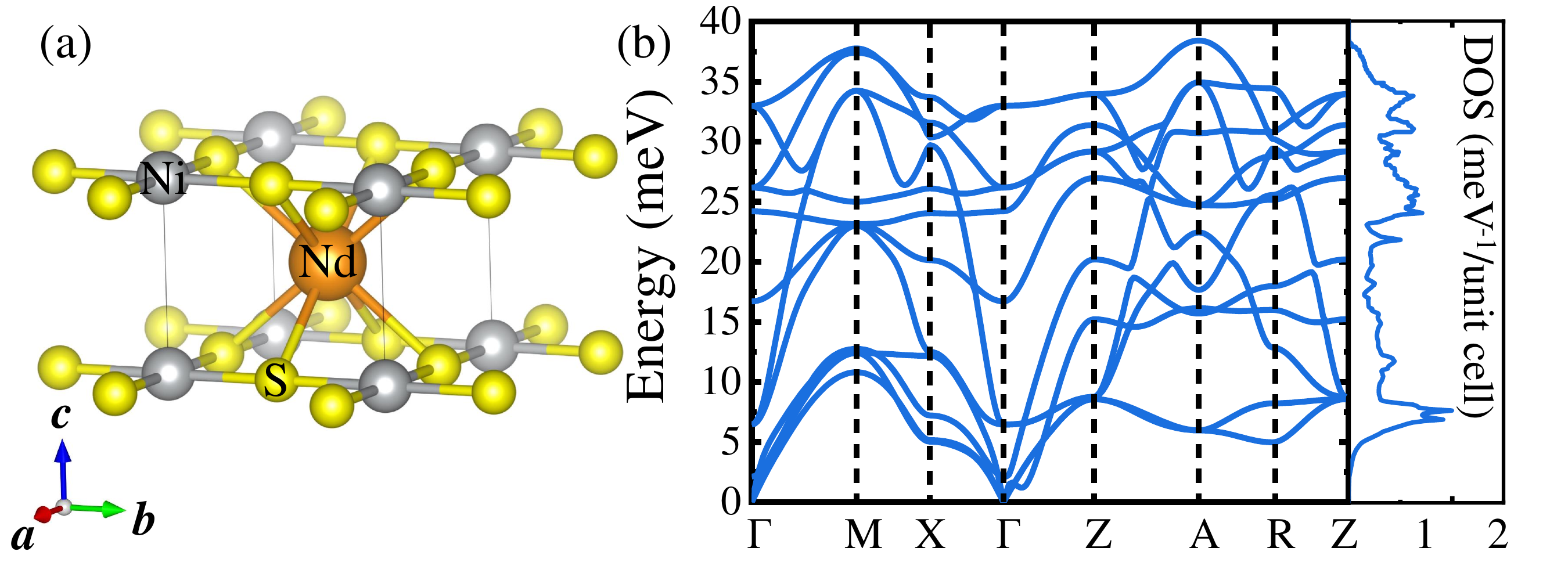}
	\vspace{-0.5cm}
	\caption{(a) Crystal structure of NdNiS$_2$, where grey, orange and yellow balls represent the Ni, Nd and S atoms respectively. (b) Phonon dispersion and corresponding density of states of NdNiS$_2$. The positivity of all phonon frequencies confirms the stability of the crystal structure.
	}
	\vspace{-0.9cm}
	\label{phonon}
	\end{center}
\end{figure}

Taking NdNiS$_2$ as an example, we first examine the stability of this compound under the same crystal structure [c.f. Fig.~\ref{phonon}(a)] as the nickelates.
Our structure optimization calculation~\cite{supp1,Baroni2001,Gonze,Hamann,Blaha2020} gives lattice constants $a=b=4.505${\AA} and $c= 3.703${\AA}.
Compared with typical structures found in other 112 materials, for example, space group 166 (LiCoO$_2$) and space group 194 (NaCoO$_2$), this structure has stronger cohesive energy by 0.7 and 1.0~eV per chemical formula, respectively, and has the lowest total energy in all the tested structures~\cite{supp1}.
Based on this optimized structure, further phonon calculation  gives fully positive frequencies, as shown in Fig.~\ref{phonon}(b).
This confirms a stable structure realizable in the laboratory.

\begin{figure}
	\begin{center}
    \includegraphics[width=7.8cm]{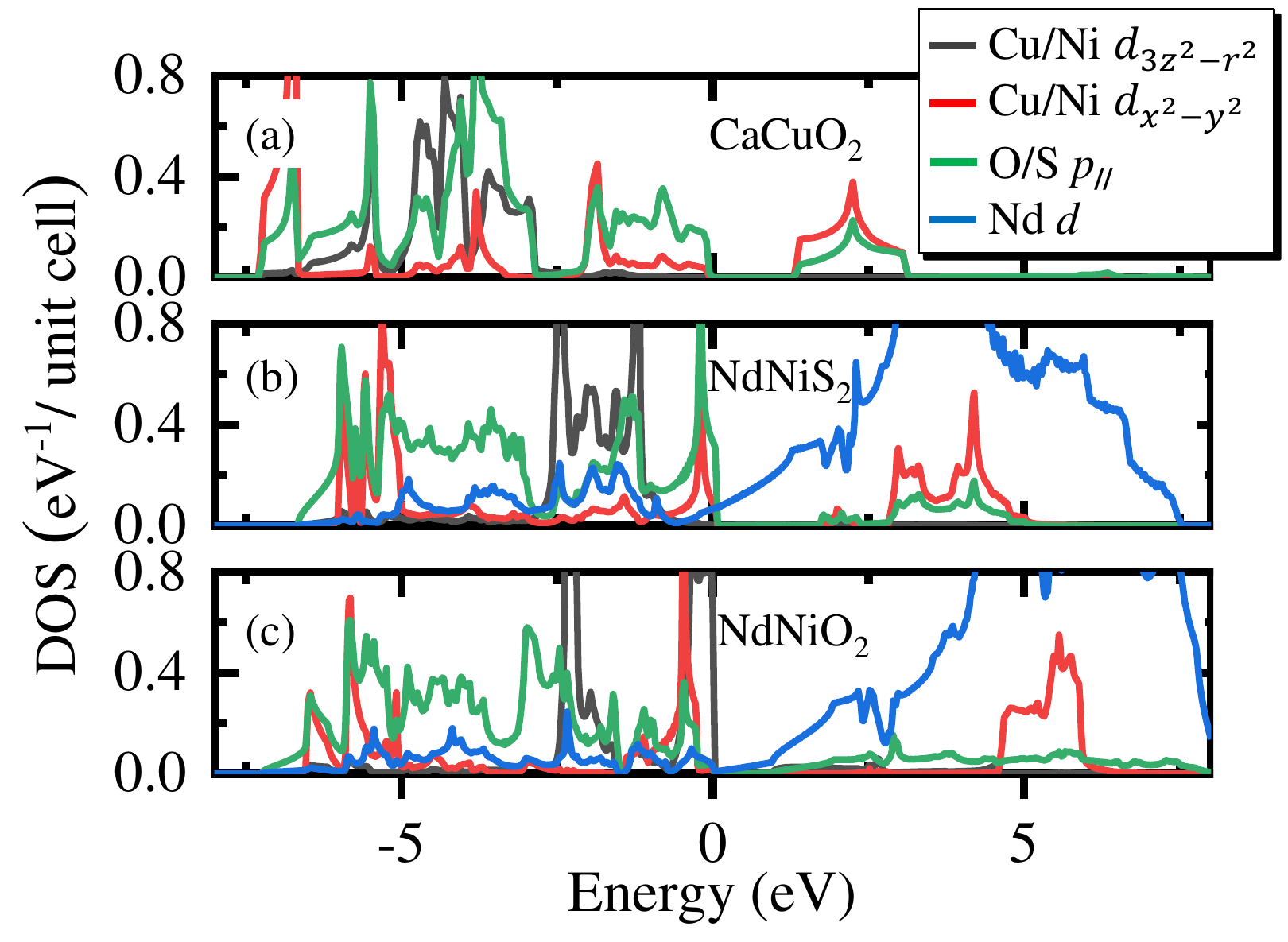}
	\end{center}
	\vspace{-0.7cm}
	\caption{Comparison of orbital-resolved Density of States in CaCuO$_2$, NdNiS$_2$, and NdNiO$_2$. Notice the gradual reduction of the relative weights of O/S  (green) $p_{\parallelsum}$ orbitals against (red) Cu/Ni $d_{x^2-y^2}$ orbitals right below the Fermi energy from CaCuO$_2$ to NdNiO$_2$.}
	\label{dos}
	\vspace{-0.4cm}
\end{figure}

Next, we verify the enhanced charge-transfer characteristic of this material.
Figure~\ref{bs}(b) shows the similar unfolded band structure of AFM NdNiS$_2$.
As expected from the above chemical intuition, substituting O by S raises the energy of the $p$ orbitals (in green) quite significantly, thereby enhancing the charge-transfer nature.
The density of state (DOS) plots in Fig.~\ref{dos} illustrate a similar trend.
Right below the Fermi energy, the relative weight of the most relevant ligand $p_{\parallelsum}$ orbitals (in green) to the $d_{x^2-y^2}$ orbital (in red) grows systematically from NdNiO$_2$ to NdNiS$_2$ and CaCuO$_2$.
(Here, $p_{\parallelsum}$ refers to O/S $p$ orbitals pointing toward nearest Cu/Ni atoms.)
Indeed, substituting O with S enhances the charge-transfer nature and brings nickel chalcogenides closer to the cuprates.

To reveal the physical benefits of a stronger charge-transfer characteristic, we proceed to investigate the low-energy effective Hamiltonian of doped holes using well-established approaches for the cuprates~\cite{supp1,Sawatzky1988,Zhang,Lang2020}.
Using DFT-parametrized high-energy many-body Hamiltonian, we calculate the local many-body ground state via exact diagonalization [the so-called (LDA+$U$)+ED method~\cite{supp1,Wei2002,Wei2006,Slater,Ogata2008}.]
The ground state with a doped hole is a spin-singlet state similar to the well-known Zhang-Rice singlet~\cite{Zhang} with (self-)doped holes mostly residing in the ligand $p$ orbitals.
[In these materials, the usage of ``LDA+$U$ '' as reference in obtaining the SU(2) symmetric effective many-body Hamiltonian is quite necessary to avoid the unphysical population of open-shell Nd $f$ orbitals.  This gives a more accurate charge density and low-energy orbitals for realistic parameters of the effective Hamiltonian.]

Note that such a strong singlet formation introduces an important correction to Figs.~\ref{bs} and \ref{dos}: it pulls the energy of the $x^2-y^2$ orbital closest to the chemical potential, even beyond the $3z^2-r^2$ orbital.
This effect, however, will still respect the abovementioned trend concerning the relative energies of O/S orbitals and Cu/Ni orbitals.

Using this singlet state as the basis, the low-energy Hamiltonian of hole carriers resembles the well-known $t$-$J$ model:
(The subspace spanned by this singlet state forms the basis for the low-energy effective Hamiltonian, upon integrating out the rest of the Hilbert space perturbatively.)
\begin{equation}
    \begin{split}
        H=\sum_{ii^\prime\nu}t_{ii^\prime}\tilde{c}_{i\nu}^\dagger \tilde{c}_{i^\prime\nu} + \sum_{<i,j>}J\mathbf{S}_i\cdot\mathbf{S}_j,
     \end{split}
    \label{H_eff}
\end{equation}
where $\tilde{c}_{i\nu}^\dagger$ creates a dressed hole at site $i$ of spin $\nu$, $\mathbf{S}_{i}=\sum_{\nu,\nu^\prime}\tilde{c}^\dagger_{i\nu}\bm{\sigma}_{\nu,\nu^\prime}\tilde{c}_{i \nu^\prime}$ denotes the spin operator and $\bm{\sigma}_{\nu,\nu^\prime}$ is the vector of Pauli matrices.
Note that the Mottness of Ni orbitals is accounted for via the constraint that excludes double occupancy in association with the strong intra-atomic repulsion of Ni $d_{x^2-y^2}$ orbital.
%Note that this formula is under the double-occupancy constraint.

Table~\ref{tab1} shows our resulting nearest neighbor hopping parameters $t_{ii^\prime}$ and super-exchange parameters $J$ for the three materials~\cite{noteJ}.
Despite the larger lattice constant in NdNiS$_2$, $t_{ii^\prime}$ turns out to be similar in all three materials owing to the larger radius of S $p$ orbitals.
In contrast, $J$ is systematically enhanced from NdNiO$_2$, to NdNiS$_2$ and CaCuO$_2$.
This is because a stronger charge-transfer nature (higher $p$ orbital energy) gives a reduced charge-transfer gap $\Delta_{CT}$ (approximate energy to return an electron from the $p_{\parallelsum}$ orbital back to the Cu/Ni $d_{x^2-y^2}$ orbital.)
With the intra-atomic repulsion being roughly the same in Cu and Ni, this in turn enhances the superexchange processes ( $\propto \Delta_{CT}^{-1}$)~\cite{Ogata2008,Lang2020}.
We stress that, despite the simplicity of such an estimation (that lacks more the complete many-body screening necessary to bring the results quantitatively closer to experimental observations), the qualitative trend among these materials is robust.

The enhanced $J$ is likely very important for the superconducting properties.
It not only could lead to a stronger magnetic correlation that dominates the low-energy physical Hilbert space, but also could give rise to a larger renormalized kinetic energy~\cite{Yin2010,Dagotto}.
In other words, a larger $J$ can stretch the energy scale of all the low-energy physics, effectively producing a larger temperature scale in the phase diagram.
One can, therefore, expect that NdNiS$_2$ should have better superconducting properties than the nickelates.

An interesting feature of NdNiS$_2$ is that the possible electron-carrier density in the parent compound will increase as a result of higher-energy $p$ orbitals (c.f. Figs.~\ref{bs} and \ref{dos}).
On the one hand, since the electron carriers are shown to be nearly decoupled from the hole carriers~\cite{Lang2020} in the nickelates (and the same is found in NdNiS$_2$~\cite{supp1}), their existence should not interfere much with the hole superconductivity.
On the other hand, these weakly correlated electron carriers might introduce additional physical effects $absent$ in the cuprates (for example, strengthening the essential superconducting phase stiffness.)
Further experimental investigation of the nickel chalcogenides will prove highly illuminating.

Finally, we note that it is not just S that has a good $p$-orbital energy, Se having a similar chemical orbital energy should also be suitable in our estimation.
This opens up a wide range of tunability in material design, for example, NdNiS$_{2-x}$O$_x$ or NdNiS$_{2-x}$Se$_x$, to optimize superconducting properties, or to explore systematic trends for better physical understanding.

\begin{table} 
    \caption{Comparison of energy difference of $d_{x^2-y^2}$ and $p_{\parallelsum}$, $\Delta_{pd}=\epsilon_{p_{\parallelsum}}-\epsilon_{d_{x^2-y^2}}$; estimated charge transfer gap, $\Delta_{CT}$; hybridization between $d_{x^2-y^2}$ and $p_{\parallelsum}$ orbitals, $t_{pd}$; nearest neighbor hopping $t$ and exchange parameter $J$ in $t-J$ model and $T_c$~\cite{Li2020,Balestrino2001} for three different materials, CaCuO$_2$, NdNiS$_2$, and NdNiO$_2$.}
    \begin{ruledtabular}
       \begin{tabular}{llll|lll}
          & $\Delta_{pd}$ & $\Delta_{CT}$ & $t_{pd}$ & $t$ & $J$  & $T_c$            \\\hline
CaCuO$_2$ & 3.7                     & $\sim$ 3.5        & 1.3      & 0.3 & $\sim$ 0.3 &  $>$50K      \\
NdNiS$_2$ & 5.7                     & $\sim$ 4.0      & 1.2      & 0.3 &  $\sim$ 0.13& ?\\
NdNiO$_2$ & 8.9                       & $\sim$ 6.0        & 1.3      & 0.3 & $\sim$ 0.07   & $\sim12$K
\end{tabular}
    \end{ruledtabular}
    \label{tab1}
	\vspace{-0.4cm}
\end{table}

In conclusion, aiming to improve the superconducting properties of the newly discovered unconventional nickelate superconductors, we identify the degree of charge transfer as the key difference with the cuprates in their high-energy electronic structure.
Guided by this, we propose another family of material nickel chalcogenides as a promising candidate for improved superconducting properties.
Taking NdNiS$_2$ as an example, we find this compound stable under the desired crystal structure and thus realizable in the laboratory.
The resulting high-energy electronic structure displays the anticipated enhancement of the charge-transfer nature.
We then reveal the physical benefits of a stronger charge-transfer characteristic via derivation of a low-energy effective Hamiltonian.
The resulting Hamiltonian encapsulates a stronger superexchange spin interaction, implying a higher temperature scale for all low-energy physics, including superconductivity.
Our study paves the way to discover more nickel-based superconductors in nickel chalcogenides with improved superconducting properties, for examples, NdNiS$_{2-x}$O$_x$ and NdNiS$_{2-x}$Se$_x$.
Further experimental exploitation of the wide range of tunability through ligand substitution would likely make significant contribution to the resolution of the longstanding puzzles of high-temperature superconductivity.

\begin{acknowledgments}
This work is supported by National Natural Science Foundation of China (NSFC) No. 11674220 and No. 11745006 and Ministry of Science and Technology No. 2016YFA0300500 and No. 2016YFA0300501.
\end{acknowledgments}
\bibliography{MainTex.bib}
\end{document}